\documentclass[a4paper, 12pt]{article}

	\usepackage[utf8]{inputenc}
	\usepackage[T1]{fontenc}
	\usepackage[top=3cm, bottom=3cm, left=2cm, right=2cm]{geometry}
	\usepackage{multicol}
	\usepackage{multirow}
	\usepackage{times}
	\usepackage{setspace}
	\usepackage{graphicx}
	\usepackage{xcolor}
	\usepackage{epsf}
	\usepackage{epsfig}
	\usepackage{wrapfig}
	\usepackage{amsmath}
	\usepackage{amssymb}
	\usepackage{mathrsfs}
	\usepackage{dsfont}
	\usepackage{blkarray}
	\usepackage{verbatim}
	\usepackage{moreverb}
	\usepackage{hyperref}
    \usepackage[super]{nth}
    \usepackage{csquotes}
    
	\hypersetup{
	colorlinks=True,
	breaklinks=true,
	urlcolor= black,
	linkcolor= black,
	citecolor = black,
	plainpages=false}
    
	\usepackage[english]{babel}
    
    \newcommand{\todo}{\textcolor{black}}
    
    \makeatletter
	\renewcommand*{\@fnsymbol}[1]{\ifcase#1\or * \or 1 \or 2 \or 3 \or 4 \or 5 \else\@arabic{\numexpr#1-1\relax}\fi}
	\makeatother
    
	\title{Impact of cyber-invasive species on a large ecological network.
    }
	\author{
		Anna \textsc{Doizy}\thanks{author for correspondence (anna@biond.org)} \ \thanks{Universit\'{e} Paris-Sud, 91400 Orsay, France and AgroParisTech, 75005 Paris, France} ,
        Edmund \textsc{Barter}\thanks{Queen's Building, University Walk, Bristol, BS8 1TR, UK} ,
		Jane \textsc{Memmott}\thanks{Life Sciences Building, Tyndall Avenue, Bristol, BS8 1TQ, UK} ,\\
        Karen \textsc{Varnham}\thanks{Royal Society for the Protection of Birds, The Lodge, Sandy, Beds, SG19 2DL, UK} \ and 
        Thilo \textsc{Gross}\thanks{Merchant Venturers Building, Woodland Road, Bristol, BS8 1UB, UK}}
    \date{2018}

\begin{document}   

	\maketitle
    
	\vfill
		
	\begin{abstract}
    	As impacts of introduced species cascade through trophic levels, they can cause indirect and counter-intuitive effects.
		To investigate the impact of invasive species at the network scale, we use a generalized food web model, capable of propagating changes through networks with a series of ecologically realistic criteria. 
		Using data from a small British offshore island, we quantify the impacts of four virtual invasive species \todo{(an insectivore, a herbivore, a carnivore and an omnivore whose diet is based on a rat)} and explore which clusters of species react in similar ways.
        \todo{We find that the predictions for the impacts of invasive species are ecologically plausible, even for large networks robust predictions for the impacts of invasive species can be obtained.
        Species in the same taxonomic group are similarly impacted by a virtual invasive species. 
        However, interesting differences within a given taxonomic group can occur.}
		The results suggest that some native species may be at risk from a wider range of invasives than previously believed.
		The implications of these results for ecologists and land managers are discussed.
	\end{abstract}
    
    \vfill
    
    \paragraph{Keywords:} 
    invasive species; introduced species; food web; generalized model; large ecological network; impact
	
\section{Introduction}
Invasive species are one of the leading threats to biodiversity in the world today \cite{IUCN} and are known to impact species functioning across many trophic levels and ecological guilds.
Accurately assessing the impact of invasive species, however, is a complicated and resource-intensive process \cite{Parker}.
\todo{By their nature, ecological communities are complex and vary greatly in time. 
Even a herbivore insect can cause indirect effects on other non-plant native species: for example, a seed feeding fly, released to control a pasture weed in the USA, caused a large increase in predatory mice \cite{Pearson}. 
Species which feed on or compete with a wide range of species and taxa may therefore have larger and more widespread effects on large parts of the food web.}
Ecological networks which characterize and quantify the \todo{species (network nodes) and their interactions (links)} can provide a powerful tool for studying the impacts of introduced species.
Several authors have studied how the impacts of invasive species on ecological networks can be modelled and propagated across trophic levels \cite{Romanuk,Novak,perturbations}. 
However, the ultimate test of these methods is whether they can produce results reliable enough to inform conservation practice and policy.	

Following \cite{perturbations}, we use a generalized model \cite{generalized}, where the abundances of the species of the network change in time according to general functions, to explore the effect of alien species on an island food web.
\todo{The generalized model allows to analyse ecologically realistic network interactions and measure the addition of virtual invasive species. 
The model finely resolves the species, which leads to a system that is an order of magnitude larger than previous systems studied with this approach, allowing for a detailed analysis \cite{Novak,perturbations}.}

Our objective is to determine how the effects of \todo{each of} four virtual introduced species, the ``cyber-invasives'', \todo{ which represent stylised classes of invaders, } propagate through an ecologically realistic network.
\todo{As an example network, we use the data based on the real ecosystem of Flat Holm, a small offshore island.}

\todo{In the methods, we explain how to construct an ecological network and we demonstrate how the impact of the introduction of an invasive species on each native species can be computed. 
In the results, we apply this methodology to the Flat Holm network. 
We find that species in the same taxonomic group respond similarly to invasive species. 
However, differences within a given taxonomic group can occur. 
Then, we comment the predictions given by the model about the perturbations caused by different cyber-invasives. 
Finally, we discuss the limits of the study, the potential improvements of the mathematical model and the overall results, along with their implications on ecological network analysis.}

\section{Methods}

\begin{figure}[ht]
				\centering
				\includegraphics[width=16cm]{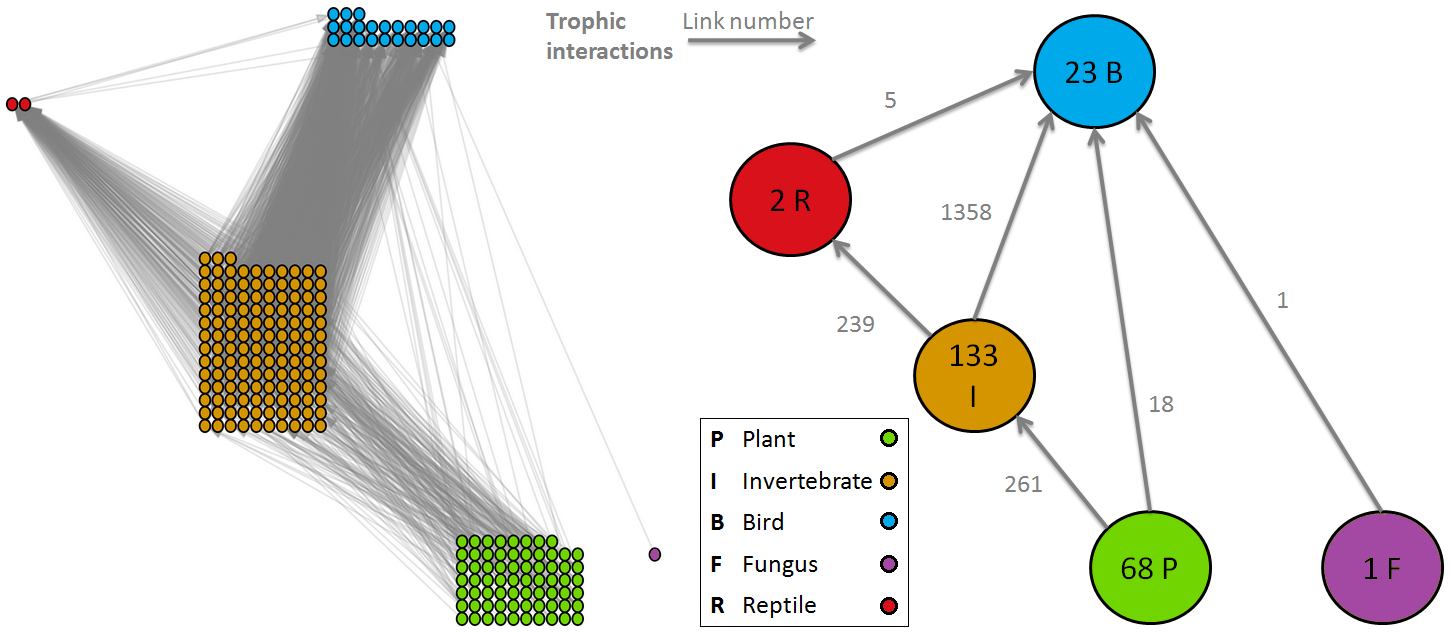}
				\caption[Representation of the Flat Holm network]{
                \small{
                The Flat Holm food web.
                Species are grouped according to their taxonomy: plant, invertebrate, bird, fungus and reptile.
                \textbf{Left:}
                The full system. 
                The dots represent the species.
                \textbf{Right:}
                The simplified system. 
                \todo{Numbers on nodes/links show the number of species/interactions that are aggregated in the formulation of the simplified system.}
                }}
				\label{representation}
			\end{figure}

Flat Holm Island is a 35ha offshore island in the Bristol Channel, UK, where very few invasive vertebrate species are present (in particular, there are no invasive non-native rats, \textit{Rattus spp.}).
We use a list of species observed on the island and data from the scientific literature to characterise their diets (supplementary information). 
This yielded a food web of 227 species: 23 birds, 2 reptiles, 133 invertebrates, 68 plants and 1 fungus (Figure \ref{representation}).
We omitted some other species because they do not trophically interact with the main food web, such as pollinators and seed-dispersers.
The system is thus significantly bigger than the the 37-species food web that was previously the largest web studied with this methodology \cite{yeakel}. 

\newpage
Generalized modelling (GM) is a universal approach to the investigation of dynamical systems \cite{generalized}. 
A generalized food web model \todo{which describes the dynamics of ecological networks, in which we consider trophic interactions as a flow of biomass was proposed in \cite{generalized}.
We define $X_i$ as the biomass density of species $i$ per unit area. 
The dynamics of each species $i$ are described by an ordinary differential equation of the form}
\begin{equation}
\label{GMODE}
\frac{\rm{d}}{\rm{d} t} X_i = G_i(X_i,T_i) + S_i(X_i) - L_i(X_i, U_i) - M_i(X_i),
\end{equation}
where the terms represent the biomass gain by predation $G$, the gain by primary production $S$, the biomass loss by predation $L$, the loss due to other causes of mortality $M$. $T_i$ and $U_i$ are the total amount of prey and predators that are effectively available to species $i$.
The key insight of generalized modelling is that the \todo{local} stability of feasible steady states \todo{(that is where the species have positive abundances in the states under consideration)} in such systems can be investigated highly efficiently without restricting the functions to specific functional forms.
The stability of steady states is governed by the system's Jacobian matrix that contains certain derivatives of Eq.~(\ref{GMODE}).

            \todo{The Jacobian matrix only contains the partial derivatives of these functions taken at the steady state, e.g. $\left. \partial G_i / \partial X_i \right|_{*}$, where $|_{*}$ means taken at the steady state. 
            Though these derivatives depend on the unknown steady state, we can assess the stability by normalizing each interaction function by the abundances taken at the steady state, e.g. $G_i^* / X_i^*$.
            \begin{equation}
			\label{Eq2}
			\left. \frac{\partial \ G_i}{\partial \ X_i} \right|_{*} = 
            \left. \frac{G_i^*}{X_i^*} 
            \frac{\partial \ \rm{log} \ G_i}{\partial \ \rm{log} \ X_i} \right|_{*}
            \end{equation}
            The right-hand side of Eq.~(\ref{Eq2}) has two parts, the first matches the scale parameters which define the biomass flows at the steady state (i.e. $G_i^* / X_i^*$ is the per capita loss rate of species $i$ due to predation at the steady state). 
            The second value, the logarithmic derivative, represents the elasticities or the exponent parameters, which measure the non-linearity of the interaction function at the steady state (for instance a value of 1 represents a linear relationship and 2 a quadratic relationship) \cite{generalized}.
            Using the same notations as in \cite{generalized} for these specific and now interpretable parameters (Tab.~\ref{parameters}), the off-diagonal elements of the \todo{Jacobian matrix \textbf{J}} can be written as}
            \begin{equation}
			\label{Eq3}
            J_{n,i} = \alpha_n \left[\rho_n \gamma_n \chi_{n,i} \Lambda_{n,i}
            - \sigma_n \left(\beta_{i,n} \psi_i 
            + \sum_{m=1}^{N} \beta_{m,n} \Lambda_{m,i} (\gamma_m -1) \chi_{m,i}
            \right) \right]
            \end{equation}
			and its diagonal elements as
            \begin{equation*}
            %\begin{align}
             J_{i,i} = \alpha_i \Biggl[ (1 - \rho_i) \phi_i 
            + \rho_i (\gamma_i \chi_{i,i} \Lambda_{i,i} + \psi_i)
            - (1 - \sigma_i) \mu_i %\\
            -  \sigma_i \Biggl(\beta_{i,i} \psi_i
            + \sum_{m=1}^{N} \beta_{m,n} \Lambda_{m,i} [(\gamma_m -1)\chi_{m,i} + 1]
            \Biggr) \Biggr]
            %\end{align}
			%\label{Eq4}
            \end{equation*}
            \todo{Intuitively, the terms in Eq.~(\ref{Eq3}) %and Eq.~(\ref{Eq4}) 
            with $\rho$ correspond to the biomass gain part of the model, i.e. $S_i$ and $G_i$ functions ($\rho_i = 0$ for primary producers and 1 for predators). 
            Terms with $\sigma$ are related to the loss in abundance of species $i$, i.e. $M_i$ and $L_i$ functions respectively, either due to natural mortality ($\mu$) or due to predation ($\sigma_i = 0$ if $i$ is a top-predator).
            The sums over $m$ take into account the feeding competition within and between species, the predation of species $m$ changes when its prey availability is modified.}

			\begin{table}[ht]
				\centering
                \footnotesize
                \begin{tabular}{cll}
                \textbf{Name} & \textbf{Interpretation} & \textbf{Rule}\\
                \hline
                &&Approximated by the reciprocal of the life span of $i$\\
                $\alpha_i$ & Rate of biomass turnover in species $i$ & assuming that the life span of birds and reptiles is 10, \\
                && plants 1, invertebrates 0.5 and fungi 0.1\\
                \hline
                $\rho_i$ & Fraction of growth in species $i$ & $\rho_i = 0$ for primary producers and $\rho_i = 1$ for predators\\
                \hline
                $\sigma_i$ & Fraction of mortality in species $i$  & $\sigma_i = 0$ for top-predators and $\sigma_i \in [0.5, 1]$\\
                & resulting from predation & chosen under a uniform law for preys\\
                \hline
                $\beta_{i,j}$ & Contribution of predation by the predator $i$ & $\beta_{i,j} = 1 / \text{\todo{number} of predators of } j$ if $i$ eats $j$\\
                && and $\beta_{i,j} = 0$ if $i$ does not eat $j$\\
                \hline
                $\chi_{i,j}$  & Contribution of prey $j$ to the available food & $\chi_{i,j} = 1 / \text{\todo{number}  of preys of } i$ if $i$ eats $j$\\
                &&  and $\chi_{i,j} = 0$ if $i$ does not eat $j$\\
                \hline
                $\mu_i$ & Elasticity of mortality in species $i$ & If the species is a top-predator it has a quadratic mortality\\
                && rate $\mu=2$, otherwise it has a linear mortality rate $\mu=1$\\
                \hline
                $\phi_i$ & Elasticity of primary production in species $i$ & $\phi_i \in [0,1]$ chosen under a uniform law\\
                \hline
                $\psi_i$ & Elasticity of predation in species $i$ & $\psi_i \in [0.5,1.5]$ chosen under a uniform law\\
                & to its own density &\\
                \hline
                $\gamma_i$ & Elasticity of predation in species $i$ & $\gamma_i \in [0.5,1.5]$ chosen under a uniform law\\
                & to the density of its prey & \\
                \hline
                $\Lambda_{i,j}$ & Elasticity of prey switching & $\Lambda = 1$ or $\Lambda = 2$ chosen under a uniform law\\
                \hline
                \end{tabular}
				\caption[Generalized model parameters]{\small{Generalized model parameters as defined in \cite{generalized} and \cite{gross2009} and the corresponding rules for the parametrization, \todo{see supplementary information}}}
				\label{parameters}
			\end{table}

The size of the model makes it hard to determine all parameters individually. Instead, we use an ensemble approach where different realizations of the model are studied, which are generated by drawing parameters randomly from plausible ranges, identified by biological reasoning.
Following \cite{perturbations}, we choose the values of the parameters such that the system exhibits realistic scaling of biomass turnover and non-linear functional responses and prey switching behaviour. 
To stabilize the relatively large system to the point where stable steady states occur, we assume quadratic mortality of top predators (supplementary information).

% It has been pointed out in \cite{Novak,perturbations} that the effect of sufficiently small perturbations on a given steady state can be estimated using the implicit function theorem which yields 
% \begin{equation}
% \boldsymbol{I} = - \textbf{J}^{-1} \boldsymbol{K}
% \label{impacteq}
% \end{equation}
% where $\boldsymbol{I}$ is a vector of indirect impacts, \todo{\textbf{J} is the Jacobian matrix defined as above} and $\boldsymbol{K}$ is a vector of the direct effects of the invasive species. 

            \todo{The direct effect of a cyber-invasive $j$ on a native species $i$ is recorded in a perturbation matrix $\textbf{K}$, defined as
            \begin{equation}
			\label{Eq5}
            K_{i,j} := \left. \frac{\partial A_i}
            {\partial Y_j} \right|_{*}
            \end{equation}
            where $A_i(X) = {\rm d}X_i /{\rm d} t$ is the right-hand side of the ODE system and $Y_j$ is the abundance of an additional population $j$ \cite{perturbations}.
            The perturbation matrix specify the direct effect on the prey of the cyber-invasive but not the relative indirect effects on the density of all species.
            To estimate these indirect effects we compute the impact matrix $\textbf{I}$, given by
            \begin{equation}
			\label{Eq6}
            I_{i,j} := \left. \frac{\partial X_i^*}{\partial Y_j^*} \right|_0
            \end{equation}
            where the derivative is evaluated in the limit of vanishing abundance $Y_j$ of the invasive species that enters the system ($|_0$).
            As long as the perturbation is small, we can make a linearity approximation and write
            \begin{equation}
			\label{Eq7}
            \left. \frac{\partial X_i^*}{\partial Y_j^*} \right|_0 = 
            \left. \frac{\partial X_i}
            {\partial A_i} \right|_{*} \left. \frac{\partial A_i}
            {\partial Y_j} \right|_{*}
            \end{equation}
            that is 
			\begin{equation}
			\textbf{I} = - \textbf{J}^{-1} \textbf{K}
			\label{impacteq}
			\end{equation}
            where $\textbf{J}^{-1}$ is the inverse of the Jacobian matrix (this is a corollary of the implicit function theorem \cite{implicit}).
            Although the approximation neglects higher order effects of the non-linearity, basic effects are captured in the Jacobian matrix. 
            The result obtained below and in previous papers \cite{yeakel} suggest that even (or perhaps especially) in large systems reliable qualitative predictions can be made in this way.}
            
            \todo{The perturbation matrix provides information about the direct effect of the perturbation; whereas the impact matrix provides information about the indirect effect of the perturbation, once it has spread through all of the network.
The impacts are quantified as the indirect relative loss or gain of biomass in the steady state, normalized with respect to the unperturbed steady state (before the arrival of the invasive) per unit of direct effect. 
These numbers should be read as proxy values with negative number indicating losses and greater absolute values indicating stronger impacts.
            %When the abundance of one species is decreased externally, negative changes propagate to its predator and positive changes propagate to its prey species.
			Thus, invasive species in this model do not cause extinctions, but instead, cause knock-on effects on the abundance of some species in the network: the model tells us the direction and relative magnitude of the perturbation.}

We use the generalized model to numerically generate an ensemble of $10^6$ steady states for the Flat Holm food web. 
For this purpose we draw the generalized model parameters randomly from the identified ranges. 
The generalized modelling procedure guarantees that all of these states are feasible in the sense that for each state we can write a plausible food web model such all species are stationary and have positive biomass densities.
We examine the Jacobian matrices of all steady states in the ensemble to determine their stability and discard all unstable states, which reduces the number of states in the ensemble to $\approx 9000$. 
For these stable steady states, we then use Eq.~(\ref{impacteq}) to estimate the impact of the four cyber-invasives, where $\textbf{K}$ is chosen to reflect the characteristics of each invasive: we study the effects of a cyber-insectivore, which feeds on all invertebrates (behaving like a hedgehog or a large shrew), a cyber-herbivore, which feeds on all plants (simulating a highly generalist vertebrate herbivore, such as a goat or rabbit), a cyber-carnivore, which feeds on all vertebrates (behaving as a generalist predator such as a cat), and \todo{a cyber-omnivore, which feeds on a subset of plants, invertebrates, birds and reptiles.
The diet of the cyber-omnivore is based on the rat \textit{Rattus rattus} and \textit{Rattus norvegicus}. Rats are very common invaders on islands and known to cause significant impacts across a wide range of taxa.
            We inferred the likely direct impact of rats from the literature (e.g. Twigg 1975 \cite{twigg}, Atkinson 1985 \cite{atkinson} and Towns et al. 2006 \cite{towns}) and from field knowledge of prey taxa likely to appeal to rats (Varnham et al., unpublished data).}

For comparison we also used the same methodology to analyse a simpler system where each group of species (birds, plants, etc.) is represented by a single network node (Fig. \ref{representation} Right, supplementary information). 
\todo{The perturbation matrix of the full system (227 species) can be found on line \cite{biond}.
For the simpler system, assuming that each cyber-invasive consumes equally all of its prey species, the perturbation matrix is
            \small
            \begin{equation}
			\label{Eq7}
            \textbf{K} = 
			\begin{blockarray}{crrrrrc}
					& P & I & B & F & R \\ 
				 \begin{block}{c(rrrrrc)}
						Carnivore & 0  & 0  & -0.1 & 0 & -0.1 &\\
						Herbivore & -0.1 & 0  & 0 & -0.1 & 0 &\\
						Insectivore & 0 & -0.1 & 0 & 0 & 0 &\\
						Rat & -0.1  & -0.1 & -0.1 & 0  & -0.1 &\\
				 \end{block}
			\end{blockarray}
			\end{equation}
            \normalsize
            where $-0.1$ means a decrease by 10\% in the targeted prey biomass.}

    \section{Results}
	\label{results_section}	
        	
\begin{figure}[ht]
	\centering
	\includegraphics[width=17cm]{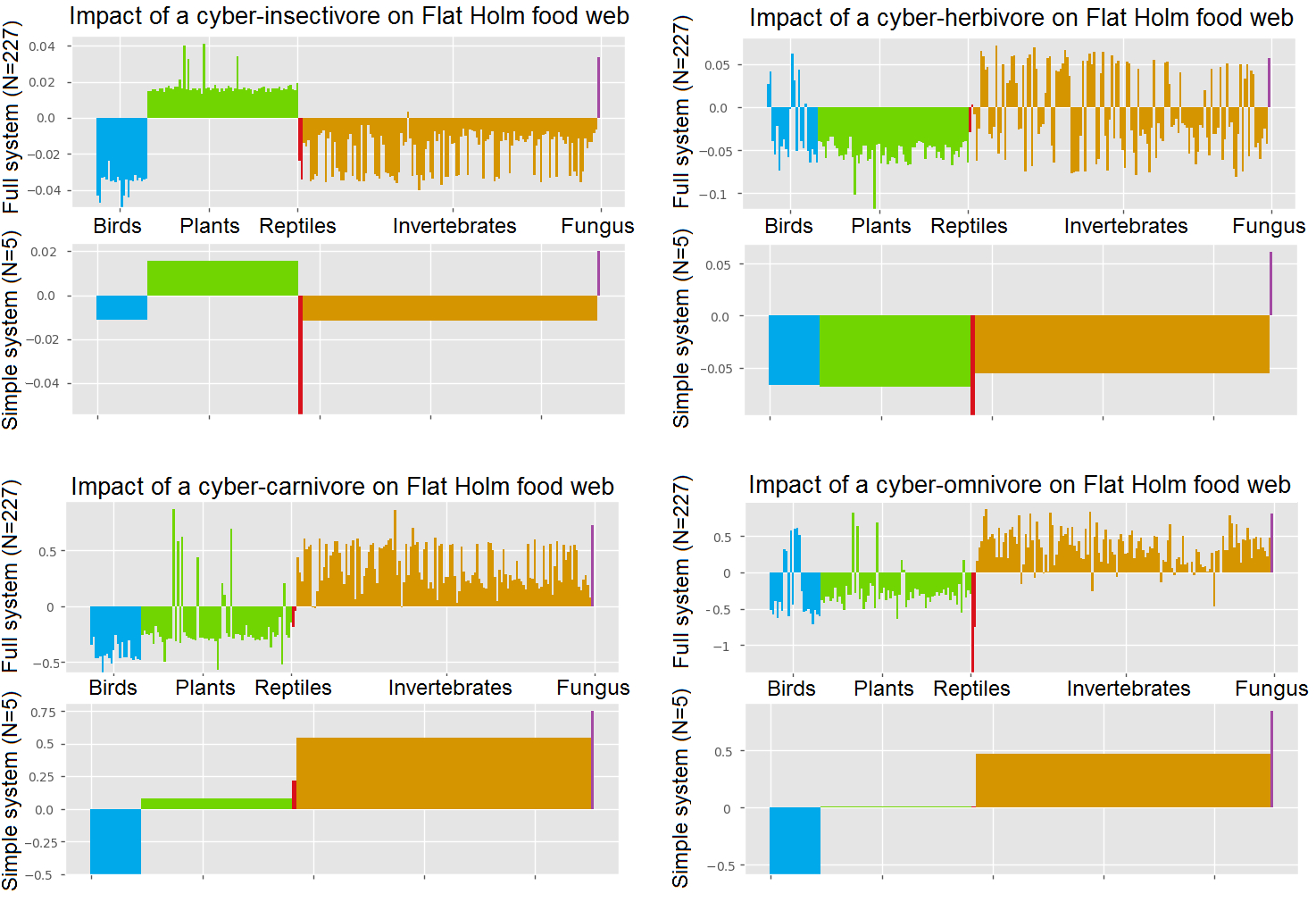}
	\caption[Impact of the cyber-invasives on both networks]{\small{Impact of the four cyber-invasives on both networks.
    The y-axis is the positive or negative impact of the invasive species.
    Each bar is one native species/taxonomic group.
    Species are grouped in five taxonomic groups: birds (blue), plants (green), reptiles (red), invertebrates (orange) and fungus (\todo{purple}).}}
	\label{Impact}
	\end{figure}

    	Considering the simple 5-species model first, the method predicts that the directly affected species and their predator are negatively impacted, while the prey and competitors of directly affected species are positively impacted (Fig. \ref{Impact}).
            These results are consistent with ecological expectations.
            
            Comparing the predictions of the 5-species model \todo{with those of} the larger network, the general pattern of impact is similar.
            However, the full model reveals greater detail with some species experiencing an impact that is the opposite than others in the same group.
            The method thus has the potential to explain patterns observed in the real world, or highlight particular species as sensitive indicators of invasive impact.
            
            In case of the cyber-herbivore the analysis reveals some issues in the underlying data. 
            The model does not resolve the resources for all of the invertebrate species. 
            In the generalized model, these species are therefore assigned an abstract carrying capacity which is not affected by the herbivore. 
            The introduction of the herbivore then benefits these species by reducing predation pressure (Fig. \ref{Impact}). 
            We verified that those invertebrate species for which the model predicts a positive impact from the cyber-herbivore are indeed those who were not assigned links to a plant resource. 
            We therefore do not expect these positive impacts to occur in the real world unless an invertebrate indeed utilizes a resource that is not negatively affected by the herbivore.
            
            The value of the detailed model is revealed in the analysis of the cyber-carnivore. 
            While the 5-species model predicts a mild positive impact on plants, the detailed model shows that a few plants are very strongly positively affected while the majority experiences a moderate negative impact (Fig. \ref{Impact}).
            For the reptiles both models predict only a slight impact by the cyber-carnivore as an increase in prey abundance counteracts direct predation. However, in the 5-species model the net impact is predicted to be positive, while it is negative in the full model. 
            
            The \todo{cyber-omnivore (or invasive rat)} has the most pervasive impact of all the invasives considered, causing large responses throughout the network (Fig. \ref{Impact}). 
            The \todo{cyber-omnivore} has a strong negative effect on most birds while those birds that are not in the \todo{cyber-omnivore}'s diet are positively impacted. 
            The reduction in birds and also in reptiles leads to a positive impact on almost all herbivores, which then leads to a negative impact on plants. 

            To understand the widespread impact of the \todo{cyber-omnivore} and the cyber-carnivore we also computed the overall sensitivity and influence of species which provides a proxy values for their role in propagating generic perturbations (formal definition in \cite{perturbations}, supplementary information).
            A sensitive species is more likely to be strongly perturbed by any disturbances propagating through the network, while an influential species may propagate the perturbation to the other species it is linked to in a more efficient way.
            Birds and reptiles proved most influential, with reptiles also showing a relatively high sensitivity to perturbations.
            The generalist bird species, having more links with the rest of the network than the others, are the most influential ones (Fig. 3 in supplementary information).
            
    \section{Discussion}
            
            In summary, our results suggest that invasive species can have a wider impact on native ecological networks than previously thought, and that these effects are both direct and indirect.
            In spite of the many assumptions made for the parametrization and the high-dimensionality of the data, the simplistic model predicts the overall pattern of impacts almost as well as the full model, but loses much of the detail.
            \todo{Moreover, the results highlight the role of birds (and more generally top predators) as important transmitters of indirect effects.}
            The catastrophic impact of invasive species on bird faunas observed in our model is well documented, including the loss of at least five species from Lord Howe Island (Australia) following the introduction of the black rat \textit{Rattus rattus} \cite{atkinson}, the loss of at least ten species in Hawaii following the introduction of avian malaria \cite{vanriper} and the loss of ten of the 12 species of forest dwelling bird on Guam following the arrival of the brown tree snake \textit{Boiga irregularis} \cite{rodda}.
			The findings from the field following invasive species introductions in temperate and tropical countries (e.g. New Zealand, UK offshore islands and Antigua) match the model's outputs for birds, reptiles and mammals (\cite{atkinson,towns,harris} and references therein).
            Despite being a relatively small component of the overall system, birds play a major role as a transmitter of cascading effects to other species in the network.
            
            There are two main limitations with our approach.
            First, the methodology constructing the food webs leads to a larger proportion of bird links than other species.
            This over-representation highlights a common problem with empirically-derived food webs, where some taxa are easier to identify than others and for which more dietary data exist \cite{may}.
            Second, a major caveat of the theoretical approach is that it provides a linearised approximation and \todo{captures some aspects of non-linearity in the Jacobian matrix, but cannot accommodate higher order effects that occur far from the stationary states and may play a role in the response to strong and sudden perturbations.}
            However this is a greater concern in small simple systems than in the large food web considered here. 
            \todo{The analysis used in the manuscript has the advantage of enabling a much safer analysis than simulation based approaches due to its advantageous numerical and statistical properties. 
            Previous papers, including for instance \cite{yeakel}, have shown that this approach allows to predict sequences of extinctions based on very limited data.}
            In the absence of better data and tools, the full network method used here presents a reasonable approach to arrive at predictions which take many properties and features of the real system into account.
            
            \todo{An obvious improvement to the model considered here would be to incorporate non-trophic interactions. 
            To a certain extent these interactions are taken into account. 
            For instance indirect effects such as exploitative and apparent competition are captured. 
            Direct (non-trophic) competition or mutualism between species of the same trophic group would constitute a competition link in the full model (which is not captured), but appears as a self-limitation term in the simplified model (which we take into account).
            The good agreement between the results of the full and simplified models therefore suggests that omitting such competition links in the full model should not have a significant effect on the result. 
            Mutualistic interactions, particularly between members of different trophic groups, e.g. pollination and seed dispersal could have a significant effect and we will focus on incorporating them in future versions of the model. 
            However, we feel that these interactions that modelling these interactions naively as biomass flows could lead to unrealistic consequences, and thus more careful modelling work is necessary.}
            
        	In reality, the impact of alien species on native communities is only likely to increase \cite{seebens2017}.
        	Ecologists need more and better tools if they are to predict their impact and plan their control, and simulation models such as the one presented here could be a part of this toolbox.
            Conservationists could allocate additional resources to surveying those species which are predicted by this model to be particularly (and unexpectedly) at risk.
            While pre- and post-eradication surveys are increasingly integrated into eradication projects (e.g. \cite{courchamp}), these have historically focused on a few taxa -- usually highly visible, charismatic and easy to survey species such as birds.
            While birds are predicted to be particularly susceptible to invasive species, as these models show other species are also likely to be affected.
            Moreover, it is not just alien predators of vertebrates that can be detrimental, herbivore and insectivore can have significant indirect negative effects.
			More and better collaboration between theoretical and empirical ecologists, with the questions addressed coming from conservation practitioners, will likely provide the most rapid progress in this field.
    
    \newpage
    \small
    \paragraph{Acknowledgements}
    We thank Ian Vaughan, Thomas Timberlake and Laura Hedon for their help.
    
    \paragraph{Declarations}
    \subparagraph{Ethical approval}
    We were not required to complete an ethical assessment prior to conducting our research.
    
    \subparagraph{Data accessibility} 
    This work did not produce new research data. 
    The data on the Flat Holm Island food web that was used in this work can be downloaded from \url{http://www.biond.org/node/581}.
    
    \subparagraph{Authors' contribution}
    AD undertook the data analysis, carried out the coding work, and drafted the manuscript; 
    EB advised on the data analysis and interpretation; 
    JM provided original idea for the paper, commented on the data analysis and edited the manuscript; 
    KV collated the data from for the food web; 
    TG oversaw the study overall, coordinated the analysis and edited the manuscript.
    All authors gave final approval for publication and agree to be held accountable for the work performed therein.
    
    \subparagraph{Conflict of interests}
    We declare no competing interests.
    
    \subparagraph{Funding}
    EB and TG were supported by EPSRC grant EP/K031686/1 and DFG research unit 1748, AD was supported by the Universit\'{e} Paris-Saclay grant and a NERC CASE PhD studentship NE/F013353/1 was awarded to KV.

	\footnotesize

	\addcontentsline{toc}{section}{References} 
	\bibliographystyle{unsrt}
	%\bibliography{biblio}
    %\printbibliography

\end{document}